\documentclass[preprint,showpacs,preprintnumbers,superscriptaddress,aps,nofootinbib]{revtex4-1}
\usepackage{amssymb}
\usepackage{amsmath,amssymb,graphicx}
\usepackage{graphicx}
\usepackage{dcolumn}
\usepackage{bm}

\usepackage{pstricks}
\usepackage{graphicx}
\usepackage{amsbsy}
\usepackage{color}
\usepackage{epsfig}
\usepackage[colorlinks=true,linkcolor=blue,citecolor=red,urlcolor=green,filecolor=blue]{hyperref}
\usepackage{slashed}
\usepackage{multirow}

\def\beq{\begin{equation}}
\def\eeq{\end{equation}}
\def\bea{\begin{eqnarray}}
\def\eea{\end{eqnarray}}

\begin{document}

\bigskip

\vspace{2cm}
\title{Analysis of the nonleptonic charmonium modes $B_s^0 \to J/\psi f_2^\prime(1525)$ and $B_s^0 \to J/\psi K^+K^-$}
\vskip 6ex
\author{C\'{e}sar A. Morales}
\email{camoralesr@ut.edu.co}
\affiliation{Departamento de F\'{i}sica, Universidad del Tolima, C\'{o}digo Postal 730006299, Ibagu\'{e}, Colombia}
\author{N\'{e}stor Quintero}
\email{nestor.quintero01@usc.edu.co}
\affiliation{Universidad Santiago de Cali, Facultad de Ciencias B\'{a}sicas, Campus Pampalinda, Calle
5 No. 62-00, C\'{o}digo Postal 76001, Santiago de Cali, Colombia}
\author{Carlos E. Vera}
\email{cvera@ut.edu.co}
\affiliation{Departamento de F\'{i}sica, Universidad del Tolima, C\'{o}digo Postal 730006299, Ibagu\'{e}, Colombia}
\author{Alexis Villalba}
\email{javillalbao@ut.edu.co}
\affiliation{Departamento de F\'{i}sica, Universidad del Tolima, C\'{o}digo Postal 730006299, Ibagu\'{e}, Colombia}

\bigskip
\begin{abstract}
In this work, we present an analysis of the nonleptonic charmonium modes $B_s^0 \to J / \psi f_2^\prime(1525)$ and $B_s^0 \to J/\psi K^+K^-$. Within the framework of the factorization approach and using the perturbative QCD for the evaluation of the relevant form factors, we find a branching fraction for  the two-body channel of BR$(B_s^{0} \to J/\psi f_2^{\prime}(1525)) = (1.6 {}^{+0.9}_{-0.7})\times 10^{-4}$ which is in agreement with the experimental values reported by the LHCb and Belle Collaborations.  The associated polarization fractions to this vector-tensor mode  are also presented. On the other hand, non-resonant and resonant contributions to the three-body decay  $B_s^0 \to J/\psi K^+K^-$ are carefully investigated. The dominant contributions of the resonances $ \phi(1020)$ and  $f_2^{\prime}(1525)$ are properly taken into account. A detailed analysis of the $K^+ K^-$  invariant mass distributions and Dalitz plot are also performed. The overall result BR$(B_s^{0} \to J/\psi K^+ K^-) = (9.3 {}^{+1.3}_{-1.1})\times 10^{-4}$ is also in satisfactory agreement with the experimental information reported by LHCb and Belle.
\end{abstract}

\maketitle
\bigskip

\section{Introduction}

The study of exclusive semileptonic and nonleptonic decays of heavy mesons $B$ and $B_s$ has provided a precise and consistent picture of the flavor sector of the Standard Model (SM) over the past decade~\cite{Bphysics}.
Some of these channels offer methods for the analysis of CP violation and determination of the angles of the unitarity triangle, test some QCD-motivated models, and the study of possible effects of physics beyond
SM~\cite{Bphysics}.
Among the possibilities of nonleptonic $B$ and $B_s$ decay modes, the color-suppressed (but CKM favored) modes induced by quark level transitions $b \to c\bar{c}s$ that involve a charmonium meson in final state are of particular interest. Specially, the charmonium vector meson $J/\psi$ is of great experimental interest because of its clean signal reconstruction ($J/\psi \to \mu^+\mu^-$)~\cite{Bphysics}. This is the case of the vector-vector mode $B \to J/\psi K^{\ast}(892)$ where the phase $\beta$, $B^0 - \bar{B}^0$ mixing parameter, can be extracted~\cite{Bphysics}. 
On the other hand, the counterpart in the $B_s$ meson system, the $B_s^{0} \to J/\psi \phi(1020)$ decay,  it is the most sensitive probe to measure the complex phase $\beta_s$ associated with the $B_s^0 - \bar{B}_s^0$ mixing process, which is extracted from the angular analysis of the time-dependent differential decay width~\cite{PDG}. Very recently,  the charmonium resonance $\psi(2S)$ has been studied in the time-dependent angular analysis of the $B_s^{0} \to \psi(2S) \phi(1020)$ decay reported by the LHCb Collaboration~\cite{LHCb:2016}.

Another interesting charmonium mode that has been studied lately by different experiments is the three-body mode $B_s^{0} \to J/\psi K^+ K^-$. It is well known that the large contribution to the  $K^+ K^-$ invariant mass spectrum of this channel is given by the vector resonance  $\phi(1020)$; i.e., the $B_s^{0} \to J/\psi K^+ K^-$ decay proceeds predominantly via  $B_s^{0} \to J/\psi \phi(1020)$~\cite{PDG}. Recently, for higher $K^+ K^-$ mass range, the significant signal of the tensor meson $f_2^{\prime}(1525)$ in the decay sequence $B_s^{0} \to J/\psi f_2^{\prime}(1525) [\to K^+ K^-]$ observed by the D0 experiment~\cite{D0:2012} has confirmed the earlier LHCb observation~\cite{LHCb:2012}. The absolute branching fractions of the mode $B_s^{0} \to J/\psi f_2^{\prime}(1525)$ and the entire mode $B_s^{0} \to J/\psi K^+ K^-$ (including resonant and non-resonant contributions) were first measured by the LHCb~\cite{LHCb:2013} and later confirmed by Belle~\cite{Belle:2013} (see Table~\ref{ExpResults}). Both measurements are in agreement with each other.
Moreover, the $B_s^{0} \to J/\psi K^+ K^-$ mode has been used to measure the CP violation parameter of the $B_s$ mixing in the $K^+ K^-$ mass region of $\phi(1020)$ resonance~\cite{LHCb:2015}. It is possible that the presence of additional resonances [with a different spin structure such as resonance $f_2^{\prime}(1525)$] to $\phi(1020)$  might affect the CP measurements~\cite{Stone:2013}. This could open new opportunities for complementary information on the parameters of CP violation~\cite{Stone:2013}. 

\begin{table}[b!]
\centering
\renewcommand{\arraystretch}{1.2}
\renewcommand{\arrayrulewidth}{0.8pt}
\caption{\small Branching fractions ($\times 10^{-4}$) of $B_s^{0} \to J/\psi f_2^{\prime}(1525)$ and $B_s^{0} \to J/\psi K^+ K^-$. For simplicity, the systematic, statistical and additional uncertainties have been combined in quadrature.}
\begin{tabular}{ccc}
\hline\hline
Mode & LHCb~\cite{LHCb:2013} & Belle~\cite{Belle:2013} \\
\hline
$B_s^{0} \to J/\psi f_2^{\prime}(1525)$ & $2.61{}^{+0.60}_{-0.54}$ & $2.60 \pm 0.81$ \\
$B_s^{0} \to J/\psi K^+ K^-$ & $7.70 \pm 0.72$ & $10.1 \pm 2.25$ \\
\hline\hline
\end{tabular} \label{ExpResults}
\end{table}

Motivated by the phenomenological importance of nonleptonic charmonium $B_s$ decays, in this work we will carry out an analysis of the modes $B_s^{0} \to J/\psi f_2^{\prime}(1525)$ and $B_s^{0} \to J/\psi K^+ K^-$. We first study the branching ratio and polarization fractions of the two-body vector-tensor mode $B_s^{0} \to J/\psi f_2^{\prime}(1525)$  and for the sake of completeness the vector-vector mode $B_s^{0} \to J/\psi \phi(1020)$ is also discussed. After that, a reanalysis of the non-resonant and resonant contributions to the $B_s^0 \to J/\psi K^+K^-$ decay is presented, where the contributions of the resonances $ \phi(1020)$ and  $f_2^{\prime}(1525)$ are properly taken into account by means of the Breit-Wigner resonance formalism. Although this mode has been previously considered in Ref.~\cite{Mohammadi}, there are some important points that have been overlooked and a more detailed analysis of the $K^+ K^-$  invariant mass distributions and Dalitz plot  will be provided in the present study. So far, it is known that there is no satisfactory treatment of nonleptonic $B_s$ to charmonium decays at present~\cite{Colangelo:2011}. Keeping this in mind, the factorization approach is used for the description of the nonleptonic charmonium $B_s$ decays under study. We will show that our results reproduce fairly well the experimental data. 

This work is organized as follows: in Sec.~\ref{BstoJpsiV}, the $B_s^{0} \to J/\psi \phi(1020)$ mode is briefly reviewed. In Sec.~\ref{BstoJpsiT}, we study the branching ratio and polarization fractions of the $B_s^{0} \to J/\psi f_2^{\prime} (1525)$ mode. The non-resonant and resonant contributions to the three-body decay  $B_s^0 \to J/\psi K^+K^-$ are carefully investigated in Sec.~\ref{3body}. Our conclusions are left for Sec.~\ref{conclusions}.

\section{$B_s^0 \to J/\psi V$ decay}  \label{BstoJpsiV}

The nonleptonic decay mode $B_s^0 \to J/\psi V$, with $V=\phi(1020)$, has been widely considered in previous works (see for instance~\cite{Colangelo:2011}). We briefly discuss its amplitude, which is written in a form that is convenient to compare with the $B_s^0 \to J/\psi f_2^{\prime}(1525)$ channel, in Sec.~\ref{BstoJpsiT}. This notation will be also helpful for discussion in Sec.~\ref{3body} where these amplitudes will be required. For the sake of completeness, the numerical result for the branching fraction is also obtained.

The  effective weak Hamiltonian ($\mathcal{H}_{eff}$) for nonleptonic charmonium $B_s$ decays induced by the $b \to c\bar{c}s$ transition is~\cite{Buras:96}
\begin{eqnarray}
\mathcal{H}_{eff} &=& \frac{G_F}{\sqrt{2}}\Big[V_{cb}V^*_{cs} (C_1
O	_1 + C_2 O_2 ) 	- V_{tb}V^*_{ts}\Big(\sum^{10}_{i=3}C_i O_i \Big) \Big] + h.c.\ ,
\end{eqnarray}

\noindent where $G_F$ is the Fermi constant, $C_i$ are the Wilson coefficients evaluated at the renormalization scale $\mu=m_b$, and $V_{ij}$  is  the respective Cabibbo-Kobayashi-Maskawa (CKM) matrix element. The four-quark local operators $O_i$ are defined as:  $O_{1-2}$ current-current (tree), $O_{3-6}$ QCD penguin, and $O_{7-10}$ electroweak penguin~\cite{Buras:96}. In Table~\ref{WilsonCoeff} we list the next to leading order (NLO) Wilson coefficients evaluated  at $\mu=m_b$~\cite{Cheng:2001}.

\begin{table}[t!]
\centering
\renewcommand{\arraystretch}{1.2}
\renewcommand{\arrayrulewidth}{0.8pt}
\caption{\small Next-to-leading Wilson coefficients evaluated at $\mu=m_b$~\cite{Cheng:2001}, where $\alpha$ is the fine-structure constant.}
\begin{tabular}{cccccccccc}
\hline\hline
$C_1$ & $C_2$ & $C_3$ & $C_4$ & $C_5$ & $C_6$ & $C_{7}/\alpha$ & $C_{8}/\alpha$ & $C_{9}/\alpha$ & $C_{10}/\alpha$ \\
\hline
1.082 & -0.185 & 0.014 & -0.035 & 0.009 & -0.041 & -0.002 & 0.054 & -1.292 & 0.263 \\
\hline\hline
\end{tabular} \label{WilsonCoeff}
\end{table}

Under the scheme of factorization, the decay amplitude of $B_s^0 \to J/\psi V$ is given by~\cite{Colangelo:2011}
\beq \label{Ampli_V}
\mathcal{M}(B_s^0 \to J/\psi V) = \dfrac{G_F}{\sqrt{2}} V_{cb} V_{cs}^{*} \  \tilde{a}_{\rm eff}  
X^{(B_sV,J/\psi)}  ,
\eeq

\noindent where using the approximation $V_{tb} V_{ts}^{*} \approx - V_{cb} V_{cs}^{*}$ (i.e. ignoring the small product $V_{ub} V_{us}^{*}$), the effective coefficient $\tilde{a}_{\rm eff}(\mu) = a_2(\mu) + a_3(\mu) + a_5(\mu)+ a_7(\mu) + a_9(\mu)$ sums the contributions from both the tree $a_{2} = C_2 + C_1/3$ and penguin $a_{2i-1} = C_{2i-1} + C_{2i}/3$ ($i= 2,3,4,5$) operators. The factorized term $X^{(B_s V,J/\psi)}$ is given by the expression
\beq \label{X_BV}
X^{(B_s V,J/\psi)} \equiv \langle J/\psi|\bar{c}\gamma_\mu c| 0\rangle\langle V |(\bar{s}b)_{V-A}| B_s \rangle \ ,
\eeq

\noindent where the hadronic matrix element $\langle J/\psi|\bar{c}\gamma_\mu c| 0\rangle = m_{J/\psi}f_{J/\psi} \epsilon^{\mu}_{J/\psi}$, with $\epsilon_{J/\psi}$ and $f_{J/\psi}$ ($m_{J/\psi}$) the vector polarization and decay constant (mass) of the $J/\psi$ meson, respectively. 
The parametrization of the $B_s \to V$ form factors can be written as~\cite{Colangelo:2011}
\begin{eqnarray}
\langle V(p_V,\epsilon_V) |\bar{s}\gamma_\mu b |B_s(P)\rangle &=& -i \dfrac{2 V^{B_s V}(q^2)}{(m_{B_s} + m_V)} \varepsilon_{\mu\nu\rho\sigma} \epsilon^{\ast\nu}_{V} P^{\rho} p_{V}^{\sigma} , \label{BstoV_V}\\
  \langle V(p_V,\epsilon_V) |\bar{s}\gamma_\mu\gamma_5 b|B_s(P)\rangle &=& 2 m_V A^{B_s V}_0(q^2) \dfrac{(\epsilon^{\ast}_{V}.P)}{q^2} q_\mu + (m_{B_s} + m_V) A^{B_s V}_1(q^2) \Big [\epsilon^{\ast}_{V\mu} - \dfrac{(\epsilon^{\ast}_{V}.P)}{q^2} q_\mu \Big] \nonumber \\
&& - A^{B_s V}_2(q^2)  \dfrac{(\epsilon^{\ast}_{V}.P)}{(m_{B_s} + m_V)}  \Big[(P+p_{V})_{\mu} - \dfrac{(m_{B_s}^2 -m_V^2)}{q^2} q_\mu \Big], \label{BstoV_A}
\end{eqnarray}

\noindent with $q_\mu = (P-p_{V})_{\mu}$ and $V^{B_s V}, A^{B_s V}_{0,1,2}$ the form factors associated with the $B_s \to V$ transition evaluated at $q^2=m_{J/\psi}^2$.  

Taking the expression of the  decay width $\Gamma(B_s^{0} \to J/\psi V)$ from~\cite{Colangelo:2011} and using the following input values: form factors obtained in the light-cone sume rules (LCSR) model~\cite{LCSR:2005}, $f_{J/\psi}=$ (416.3 $\pm$ 5.3) MeV~\cite{Colangelo:2011}, NLO Wilson coefficients evaluated  at $\mu=m_b$ (Table \ref{WilsonCoeff}), CKM matrix elements $|V_{cb}| = (41.1 \pm 1.3)\times 10^{-3}$, $|V_{cs}|= 0.986 \pm 0.016$, $\tau_{B_s} = 1.510 \times 10^{-12} \ s$ and masses of the mesons~\cite{PDG}; we get a value of 
\begin{equation} \label{BR:VV}
{\rm BR}(B_s^{0} \to J/\psi \phi(1020)) = (10.4 \pm 0.3) \times 10^{-4},
\end{equation}

\noindent which is consistent with the experimental value $(10.8 \pm 0.9) \times 10^{-4}$~\cite{PDG}.

\section{$B_s^0 \to J/\psi T$ decay}  \label{BstoJpsiT}

Sharing the same CKM mixing elements and penguin contributions of the $B_s^0 \to J/\psi V$ mode, the decay amplitude of $B_s^0 \to J/\psi T$ [with $T=f_2^\prime(1525)$] is written as
\beq \label{Ampli_T}
\mathcal{A}(B_s^0 \to J/\psi T) = \dfrac{G_F}{\sqrt{2}} V_{cb} V_{cs}^{*} \ \tilde{a}_{\rm eff}  X^{(B_sT,J/\psi)}  ,
\eeq

\noindent where the factorized term $X^{(B_sT,J/\psi)}$ has the expression
\beq \label{X_BT}
X^{(BT,J/\psi)} \equiv \langle J/\psi|\bar{c}\gamma^\mu c| 0\rangle \langle T|(\bar{s}b)_{V-A}| B_s \rangle .
\eeq

\noindent In analogy to the hadronic matrix element that describes $B_s \to V$ transition, the structure of the $B_s \to T$ form factors is the same by adequately replacing the $\epsilon^{\mu}_V$ polarization vector by a \textit{new polarization vector} $\epsilon^{\mu}_{T} = \tilde{\epsilon}^{\mu\nu} P_{\nu} /m_{B_s}$ in Eqs.~\eqref{BstoV_V} and~\eqref{BstoV_A}~\cite{Cheng:2010,Wang11}, with $\tilde{\epsilon}^{\mu\nu}$ being the polarization of the spin-2 tensor meson and $P$ the $B_s$ meson momentum (see appendix~\ref{appA} for details). 
In this case $V^{B_sT}$ and $A^{B_sT}_{0,1,2}$ are the form factors associated with the $B_s \to T$ transition. In ensuing calculations we will use the theoretical predictions provided by the perturbative QCD (pQCD) approach~\cite{Wang11}. Within the pQCD approach the $q^2$-dependence of the form factors $V^{B_sT}$ and $A^{B_sT}_{0,1}$ can be represented by the three-parameter formula~\cite{Wang11}
\begin{equation}\label{1}
F^{B_sT}(q^2) = \frac{F^{B_sT}(0)}{(1-q^2/m_{B_s}^2)(1-a q^2/m_{B_s}^2+b(q^2/m_{B_s}^2)^{2})},
\end{equation}

\noindent where the parameters $a$, $b$ and $F^{B_sT}(0)$ (value at the zero momentum transfer) for $B_s \to f_2^\prime(1525)$ transition are displayed in Table~\ref{FF} (taken from Table II of Ref.~\cite{Wang11}). While the form factor $A^{B_sT}_2$ can be expressed as a linear combination of $A^{B_sT}_0$ and $A^{B_sT}_1$~\cite{Wang11}
\beq \label{FF_A2}
A^{B_sT}_2(q^2) = \dfrac{(m_{B_s}+m_{T})}{m_{B_s}^2- q^2} [(m_{B_s}+m_{T}) A^{B_sT}_1(q^2) -2 m_{T} A^{B_sT}_0(q^2)].
\eeq

We will assume the $f_2^\prime(1525)$ meson as a $s\bar{s}$ state (since mainly $f_2^\prime(1525) \to K^+ K^-$~\cite{PDG}) and we will neglect the small mixing angle ($\sim 9^\circ$~\cite{PDG}) between the two isosinglet mesons $f_2(1270) - f_2^\prime(1525)$.

\begin{table}[t!]
\centering
\renewcommand{\arraystretch}{1.2}
\renewcommand{\arrayrulewidth}{0.8pt}
\caption{\small Form factors for $B_s^{0} \rightarrow f_{2}^{\prime}(1525)$ transitions obtained in the pQCD approach~\cite{Wang11} (uncertanties added in quadrature) are fitted to the three-parameter form Eq.~\eqref{1}.}
\begin{tabular}{cccc}
\hline\hline
$F^{B_sT}$ & $F^{B_sT}(0)$ & $a$ & $b$ \\
\hline
$V^{B_s f_{2}^{\prime}(1525)}$ &  0.20${}^{+0.06}_{-0.04}$ & 1.75${}^{+0.05}_{-0.03}$ & 0.69${}^{+0.09}_{-0.01}$ \\
$A_0^{B_s f_{2}^{\prime}(1525)}$ &  0.16${}^{+0.04}_{-0.03}$ & 1.69${}^{+0.04}_{-0.03}$ & 0.64${}^{+0.01}_{-0.04}$ \\
$A_1^{B_s f_{2}^{\prime}(1525)}$ &  0.12${}^{+0.04}_{-0.03}$ & 0.80${}^{+0.07}_{-0.03}$ & $-0.11{}^{+0.10}_{-0.00}$ \\
\hline\hline
\end{tabular} \label{FF}
\end{table}

The explicit expression for the decay width of $B_s^{0} \to J/\psi T$ has the form 
\bea \label{Gamma_BtoVT}
\Gamma(B_s^{0} \to J/\psi T) &=& \frac{G_F^2}{48\pi m_{B_s}^3} \frac{|V_{cb}V_{cs}^\ast |^2 \tilde{a}_{\rm eff}^2 f_{J/\psi}^2}{16 m_{B_s}^2 m_{T}^4} \ \Big[ \alpha_T \lambda_T^{7/2} + \beta_T \lambda_T^{5/2} + \gamma_T \lambda_T^{3/2} \Big], 
\eea

\noindent where $\lambda_T \equiv \lambda(m_{B_s}^2,m_T^2,m_{J/\psi}^2)$, with $\lambda(x,y,x) = x^2 + y^2 + z^2 - 2 (xy+xz+yz)$ the usual kinematic K\"{a}llen function, and
\bea
\alpha_T &=&  \frac{[A^{B_sT}_2(q^2)]^2}{(m_{B_s} + m_T)^2}, \\
\beta_T  &=&  \frac{6 q^2 m_{T}^2}{(m_{B_s} + m_T)^2} [V^{B_sT}(q^2)]^2 - 2 (m_{B_s}^2 - m_T^2 - q^2) A^{B_sT}_1(q^2) A^{B_sT}_2(q^2), \\
\gamma_T &=&  (m_{B_s} + m_T)^2 (\lambda_T + 10  q^2 m_T^2) [A^{B_sT}_1(q^2)]^2 . 
\eea

\noindent As it was pointed out in~\cite{Munoz}, it is worth to notice that the $\lambda_T^{L + 1/2} \propto |\vec{p}_T|^{2L+1}$ (with $|\vec{p}_T|$ being the three-momentum magnitude of the tensor meson in the $B_s$ rest frame) dependence in Eq.~\eqref{Gamma_BtoVT} indicates that  in vector-tensor modes the orbital angular momentum of the wave $L= 1$, 2, and 3 are simultaneously allowed, as expected.

Taking the same numerical input values as in Sec.~\ref{BstoJpsiV} and the form factors from the pQCD approach~\cite{Wang11} (Table~\ref{FF}), the branching ratio is found to be\footnote{Using the predictions of the form factors derived from LCSR~\cite{Yang11}, we have obtained a value ${\rm BR}(B_s^{0} \to J/\psi f_2^{\prime}(1525)) = (1.1 \pm 0.3)\times 10^{-4}$, which is smaller than~\eqref{BR:VT} and the experimental measurements (see Table~\ref{ExpResults}).} 
\begin{equation} \label{BR:VT}
{\rm BR}(B_s^{0} \to J/\psi f_2^{\prime}(1525)) = (1.6 {}^{+0.9}_{-0.7})\times 10^{-4},
\end{equation}

\noindent where the theoretical error corresponds to the uncertainties due to the CKM elements, decay constant and form factors (mainly dominated by the latter). 
Within the error bars our result is in agreement with the experimental values reported by LHCb~\cite{LHCb:2013} and Belle~\cite{Belle:2013} (see Table~\ref{ExpResults}). In comparison to previous theoretical estimation of $(3.3 \pm 0.5)\times 10^{-4}$ obtained in~\cite{Mohammadi}, our result turns out to be lower than this. 
In addition, based on the chiral unitary approach for mesons, the authors of Ref.~\cite{Oset} have been estimated the ratio of branching fractions
\beq
\frac{{\rm BR}(B_s^{0} \to J/\psi f_2(1270)) }{{\rm BR}(B_s^{0} \to J/\psi f_2^{\prime}(1525))} = (8.4 \pm 4.6)\times 10^{-2},
\eeq

\noindent that is compatible within errors	with the experiment~\cite{Oset}.  

	
Finally, as a by-product, using Eqs.~\eqref{BR:VT} and~\eqref{BR:VV} we also estimate the ratio between the vector-tensor mode $B_s^{0} \to J/\psi f_2^{\prime}(1525)$ and vector-vector mode $B_s^{0} \to J/\psi \phi(1020)$ 
\beq
R_{f_2^\prime / \phi} \equiv \frac{{\rm BR}(B_s^{0} \to J/\psi f_2^{\prime}(1525)) }{{\rm BR}(B_s^{0} \to J/\psi \phi(1020)) } = (15.4{}^{+9.0}_{-7.0}) \%,
\eeq

\noindent that is consistent with different experimental measurements $(25.0 \pm 6.0) \%$  LHCb~\cite{LHCb:2013}, $(19.0 \pm 6.0)\%$ D0~\cite{D0:2012} and $(21.5 \pm 5.5)\%$ Belle~\cite{Belle:2013} .

\subsection{Polarization fractions}
 
In this subsection we study the polarizaton fractions of the decay mode $B_s^{0} \to J/\psi T$. Taking advantage to the fact that this vector-tensor mode can be treated as the vector-vector mode $B_s \to J/\psi V$, by just replacing $\epsilon^{\mu}_V$ by $\epsilon^{\mu}_{T}$ previously introduced, the factorizable transition amplitude~\eqref{Ampli_T} can be generically decomposed in terms of the invariant amplitudes $\mathbf{a}$, $\mathbf{b}$ and $\mathbf{c}$~\cite{Angular}
\bea
\mathcal{M}(B_s^0 \to J/\psi T)  &=& \mathbf{a} (\epsilon_{J/\psi}^\ast \cdot \epsilon_T^\ast) + \frac{\mathbf{b}}{m_{J/\psi} m_T} (\epsilon_{J/\psi}^\ast \cdot P) (\epsilon_T^\ast \cdot P) \nonumber \\
&& + i  \frac{\mathbf{c}}{m_{J/\psi} m_T} \varepsilon_{\mu\nu\alpha\beta} \epsilon_T^{\ast \mu} \epsilon_{J/\psi}^{\ast \nu} p_T^\alpha P^\beta ,
\eea

\noindent where 
\bea
\mathbf{a} &=&  - \xi  (m_{B_s} + m_T) A_1^{B_sT}(m_{J/\psi}^2), \\
\mathbf{b} &=&  \xi m_{J/\psi} m_T \dfrac{2 A_2^{B_sT}(m_{J/\psi}^2)}{(m_{B_s} + m_T)},\\
\mathbf{c} &=& \xi m_{J/\psi} m_T\dfrac{2 V^{B_sT}(m_{J/\psi}^2)}{(m_{B_s} + m_T)}, 
\eea

\noindent are expressed in terms of $V^{B_sT}, A^{B_sT}_{1,2}$ and the global factor $\xi = i G_F V_{cb} V_{cs}^{*} \tilde{a}_{\rm eff} f_{J/\psi}m_{J/\psi}/\sqrt{2}$. The longitudinal ($\mathcal{H}_{0}$) and transverse ($\mathcal{H}_{\pm}$) helicity amplitudes can be expressed in terms of $\mathbf{a}$, $\mathbf{b}$ and $\mathbf{c}$ as~\cite{Chen:2007,Cheng:2010}
\bea
\mathcal{H}_{0} &=&  -\sqrt{\frac{2}{3}} \dfrac{|\vec{p}_T|}{m_T} [\mathbf{a}x + \mathbf{b} (x^2 - 1)],  \\
\mathcal{H}_{\pm} &=&  \dfrac{1}{\sqrt{2}} \dfrac{|\vec{p}_T|}{m_T} [\mathbf{a} \pm \mathbf{c}\sqrt{x^2 -1}],
\eea

\noindent with $x=(m_{B_s}^2 - m_{J/\psi}^2 - m_T^2)/2m_{J/\psi} m_T$ and $|\vec{p}_T| = \sqrt{\lambda_T} /2m_{B_s}$. In addition, the transverse amplitudes (parallel and perpendicular) defined in the transversity basis (also refer as linear polarization basis) are related to the helicity ones via~\cite{Angular}
\bea
\mathcal{A}_{0} &=&  \mathcal{H}_{0}, \nonumber \\
\mathcal{A}_{\parallel} &=& \dfrac{1}{\sqrt{2}}(\mathcal{H}_{+} + \mathcal{H}_{-}) = \dfrac{|\vec{p}_T|}{m_T} \mathbf{a}, \\   
\mathcal{A}_{\perp} &=& \dfrac{1}{\sqrt{2}}(\mathcal{H}_{+} - \mathcal{H}_{-}) = \dfrac{|\vec{p}_T|}{m_T} \mathbf{c}\sqrt{x^2 -1}. \nonumber
\eea

\noindent The decay rate can be expressed in terms of these amplitudes as~\cite{Chen:2007,Cheng:2010}
\bea
\Gamma(B_s^{0} \to J/\psi f_2^{\prime}(1525)) &=& \dfrac{\sqrt{\lambda_T}}{16\pi m_{B_s}^3} \sum_{i=0,\pm} |\mathcal{H}_{i}|^2 , \\
&=& \dfrac{\sqrt{\lambda_T}}{16\pi m_{B_s}^3} \sum_{i=0,\parallel,\perp} |\mathcal{A}_{i}|^2 . 
\eea

In terms of the transversity basis, the longitudinal and parallel (perpendicular) polarization fractions are defined as~\cite{Cheng:2010}
\bea
f_L &=& \frac{|\mathcal{A}_{0}|^2}{|\mathcal{A}_{0}|^2 + |\mathcal{A}_{\parallel}|^2 + |\mathcal{A}_{\perp}|^2}, \label{fL} \\
f_{\parallel (\perp)} &=& \frac{|\mathcal{A}_{\parallel (\perp)}|^2}{|\mathcal{A}_{0}|^2 + |\mathcal{A}_{\parallel}|^2 + |\mathcal{A}_{\perp}|^2}, \label{fperp}
\eea

\noindent respectively. The transverse polarization fraction is $f_T = (1- f_L)$. By definition the fractions \eqref{fL} and \eqref{fperp} satisfy the relation $f_L + f_\parallel + f_\perp =1$. The numerical results for the polarization fractions $f_L, f_\parallel$, and  $f_\perp$ are 
\bea
f_L (B_s^{0} \to J/\psi f_2^{\prime}(1525)) &=& (53.3 \pm 18.0)\%  , \nonumber \\
f_\parallel (B_s^{0} \to J/\psi f_2^{\prime}(1525)) &=& (30.8 \pm 12.0) \% , \\
f_\perp (B_s^{0} \to J/\psi f_2^{\prime}(1525)) &=& (15.8 \pm 0.60) \% , \nonumber
\eea

\noindent respectively. Although it is expected that vector-tensor modes will be dominated by the longitudinal polarization~\cite{Cheng:2010}, we get  within the errors the ratio $f_T/f_L(J/\psi f_2^{\prime}) \sim 1$ implying that the two fractions $f_T$ and $f_L$ are roughly equal. A similar theoretical result is obtained in the $B_s^{0} \to J/\psi \phi(1020)$ mode, i.e. $f_T/f_L( J/\psi \phi) \sim 1$~\cite{Colangelo:2011,Wang:2014}, which is in agreement with the measurement of the longitudinal polarization fraction $f_L(B_s^{0} \to J/\psi \phi(1020)) = (49.7 \pm 3.3) \%$ reported by LHCb~\cite{LHCb:2011}. In addition, our results for the polarization fractions are in accordance with the fit fractions in the helicity basis obtained by LHCb in the amplitude analysis of the $B_s^0 \to J/\psi K^+K^-$ decay for the resonance $f_2^{\prime}(1525)$~\cite{LHCb:2013}. Nevertheless, with the integrated luminosity collected by the LHCb detector during LHC Run 1 (3 fb${}^{-1}$ at $\sqrt{s} = 7$ and  8 TeV) and that expected during LHC Run 2 (additional 5 fb${}^{-1}$ at $\sqrt{s} = 14$ TeV), it will be an interesting independent measurement of the helicity components $+$ and  $-$ (or $\parallel$ and $\perp$ components) to test our results.

\section{Non-resonant and resonant contributions to $B_s^0 \to J/\psi K^+K^-$ decay} \label{3body}

The three-body charmonium mode $B_s^0 \to J/\psi K^+K^-$ receives both non-resonant and resonant contributions~\cite{LHCb:2013}. Although this channel has been previously considered in Ref.~\cite{Mohammadi},  in this section we provide a detailed reanalysis of such contributions. We also stress some important points that were overlooked by the authors of Ref.~\cite{Mohammadi}. 

In the framework of the factorization approach the decay amplitude associated with the non-resonant (NR) contribution of the $B_s^0 \to J/\psi K^+K^-$ mode has the form
\bea \label{Amp_NR}
\mathcal{M}(B_s^0 \to J/\psi K^+K^-)_{\rm NR} &=& \dfrac{G_F}{\sqrt{2}} V_{cb} V_{cs}^{*} \ \tilde{a}_{\rm eff}  \langle J/\psi|(\bar{c}c)_{V-A}| 0\rangle \nonumber \\
&& \times \langle K^+ K^- |(\bar{s}b)_{V-A}|B_s \rangle_{\rm NR}  ,
\eea

\noindent where only the current-induced process with a meson emission is present~\cite{Cheng:2002}. In the heavy meson chiral perturbation theory~\cite{Wise:1992}, the hadronic matrix element $\langle K^+ K^- |(\bar{s}b)_{V-A}|B_s \rangle_{\rm NR}$ can be written in terms of four NR form factors $r$, $w_{\pm}$, and $h$ that are defined by~\cite{Wise:1992,Cheng:2014}
\begin{eqnarray}
  \langle K^+(p^\prime) K^-(p) |(\bar{s}b)_{V-A} |B_s(P)\rangle_{\rm NR}  &=& i r (P- p - p^\prime)_\mu + i w_+ (p^\prime +p)_\mu + i w_- (p^\prime - p)_\mu \nonumber \\
&& - 2 h \varepsilon_{\mu\nu\alpha\beta} P^\nu p^{\prime\alpha} p^\beta .
\end{eqnarray}

\noindent In the present case the NR form factors $w_{\pm}$ and $h$ contribute while $r$ vanishes due to the condition $\epsilon_{J/\psi} \cdot p_{J/\psi} = 0$. These are explicitly given by the expressions~\cite{Wise:1992,Cheng:2014} 
\bea
w_+ &=& -\frac{g}{f_K^2}\dfrac{f_{B^\ast}\sqrt{m_{B^\ast}^3 m_{B_s}}}{s-m_{B^\ast}^2} \Big[1- \Big(  \dfrac{m_{B_s}^2 - m_K^2 -s}{2m_{B^\ast}^2}\Big)\Big] + \frac{f_{B_s}}{2 f_K^2}, \\
w_- &=& \frac{g}{f_K^2}\dfrac{f_{B^\ast}\sqrt{m_{B^\ast}^3 m_{B_s}}}{s-m_{B^\ast}^2} \Big[1+ \Big(  \dfrac{m_{B_s}^2 - m_K^2 -s}{2m_{B^\ast}^2}\Big)\Big], \\
h &=& \frac{2 g^2 f_{B_s}}{f_K^2} \dfrac{m_{B_s}^2}{(m_{B_s}^2 - m_{J/\psi}^2 -t) (s+m_{B_s}^2 - m_K^2)}, 
\eea

\noindent where $s \equiv m^2(J/\psi K^+) =(p_{J/\psi} + p^\prime)^2$ and $t \equiv m^2(K^+K^-)=(p^\prime +p)^2$ are the kinematical variables that represent the $J/\psi K^+$ and $ K^+ K^-$ invariant masses, respectively. The heavy-flavor independent strong coupling $g$ can be extracted from the CLEO measurement of the $D^{\ast +}$ decay width, $|g| = 0.59 \pm 0.07$~\cite{CLEO:2001}. For the pole mass and decay constants we will take the following numerical inputs: $m_{B^\ast} = 5324.83$ MeV~\cite{PDG} and $f_K = (155.6 \pm 0.4)$ MeV~\cite{Rosner:2015}, $f_{B_s} = (226.0 \pm 2.2)$ MeV~\cite{Rosner:2015}, $f_{B^\ast}= (175 \pm 6)$ MeV~\cite{HPQCD}. 

On the other hand, the resonant (R) contributions are usually described in terms of the
Breit-Wigner (BW) resonance formalism. The three-body matrix element in~\eqref{Amp_NR} is written as~\cite{Cheng:2002,Cheng:2014}
\bea
\langle K^+(p^\prime) K^-(p) |(\bar{s}b)_{V-A} |B_s(P)\rangle_{\rm R} &=& \sum_{R} {\rm BW}_{R}(t) \ g_{R K^+K^-} \ \epsilon_{R}\cdot(p^\prime - p)  	\nonumber \\ 
&& \times \langle R |(\bar{s}b)_{V-A}|B_s \rangle , \label{R_part}
\eea

\noindent where $g_{R K^+K^-}$ is the strong coupling constant and 
\beq \label{BW}
{\rm BW}_{R}(t) = \dfrac{1}{m^2_{R}- t - im_{R}\Gamma_{R}(t)} ,
\eeq

\noindent is the BW function of the intermediate resonant state $R$, with $m_{R}$ and $\Gamma_{R}(t)$ being its respective mass and decay width of $R \to K^+K^-$. We adopt the $t$-dependent parametrization for the decay width~\cite{LHCb:2013}
\beq
\Gamma_{R}(t) = \Gamma_{0R} \Big(\dfrac{m_R^2}{t} \Big) \Big[ \dfrac{Q(t)}{Q(m_R^2)}\Big]^{2 L_{R}+1} F_R^2
\eeq

\noindent where $\Gamma_{0R} $ is the resonance width at its peak and $Q(t) = \lambda(t,m_{K^+}^2,m_{K^-}^2)^{1/2}/2\sqrt{t}$ is the momentum of the $K^+$ (or the $K^-$) evaluated in the $K^+K^-$ rest frame. The orbital angular momentum is $L_R = 1 \ (2)$ for vector (tensor) and the Blatt-Weisskopf barrier factors $F_R$ are taken from~\cite{LHCb:2013}.  The sum in~\eqref{R_part} is extended over all possible resonant contributions. Although different resonances can appear (such as $f_0(980), f_0(1370), \phi(1680)$, $f_2(1750)$ and $f_2(1950)$~\cite{LHCb:2013}), we will take the intermediate vector $\phi(1020)$ and tensor $f_2^\prime(1525)$ mesons as the most important ones~\cite{LHCb:2013}. Furthermore, it was found by LHCb that the interference contributions between two different spin resonances and between NR and R components are zero~\cite{LHCb:2013} and therefore,  as a good approximation, the interference between these components will not be considered here. 


The R amplitude of $B_s^0 \to J/\psi K^+K^-$ is then given by 
\bea  - s^2
\mathcal{M}(B_s^0 \to J/\psi K^+K^-)_{\rm R} &=& i \dfrac{G_F}{\sqrt{2}} V_{cb} V_{cs}^{*} \ \tilde{a}_{\rm eff}    \sum_{R} {\rm BW}_{R}(t) \ g_{R K^+K^-}\nonumber \\
&& \times \ \epsilon_{R}\cdot(p^\prime - p)  X^{(B_s R,J/\psi)},
\eea

\noindent with $X^{(B_s R,J/\psi)}$ the factorized terms coming from $R = V$ and $T$, given by~\eqref{X_BV} and~\eqref{X_BT}, respectively.  From the decay amplitude of the strong decays $R \to K^+K^-$
\bea 
\mathcal{M}(V \to  K^+K^-) &=& g_{V K^+K^-} \ \epsilon_{V}^\mu (p^\prime - p)_\mu  , \\
\mathcal{M}(T \to  K^+K^-) &=& g_{T K^+K^-} \tilde{\epsilon}^{\mu\alpha} p^\prime_\mu p_\alpha, 
\eea

\noindent the strong coupling constants $g_{R K^+K^-}$ are determined from the experimental value of decay width of $R \to K^+K^-$ via the expressions
\bea
g_{VK^+K^-} &=& \sqrt{\dfrac{48 \pi m_V^5 \Gamma(V\to K^+K^-)}{\lambda(m_V^2,m_{K^+}^2,m_{K^-}^2)^{3/2}}} , \label{g_VKK} \\
g_{T K^+K^-} &=& \sqrt{\dfrac{1920 \pi m_{T}^7 \Gamma(T\to K^+K^-)}{\lambda(m_T^2,m_{K^+}^2,m_{K^-}^2)^{5/2}}}. \label{g_TKK}
\eea

\noindent Using the above expressions and the experimental measurements $\Gamma(\phi(1020) \to K^+K^-) = (2.08 \pm 0.04)$ MeV and $\Gamma(f_2^{\prime}(1525) \to K^+K^-) = (64.75{}^{+7.06}_{-5.93})$ MeV~\cite{PDG}, we get $g_{VK^+K^-} = 4.47 \pm 0.03$ and $g_{T K^+K^-} = 20.70{}^{+0.89}_{-0.75}$ ${\rm GeV}^{-1}$, respectively. The error reported is due to the experimental uncertainty in the decay width. 
Let us notice an important point that has been overlooked  in Ref.~\cite{Mohammadi}, since the same expression has been used to obtain $g_{RK^+K^-}$ for both $V$ and $T$, namely Eq. (30) of~\cite{Mohammadi} [Eq.~\eqref{g_VKK} of this work]. This is a mistake since~\eqref{g_VKK} only allows us to obtain the strong coupling for $V=\phi(1020)$, while~\eqref{g_TKK} allows us to obtain the one for $T=f_2^\prime(1525)$. Besides, $g_{VK^+K^-}$ is dimensionless, while $g_{T K^+K^-}$ has dimensions of ${\rm GeV}^{-1}$. Indeed, by  employing Eq. (30) of~\cite{Mohammadi} and the experimental measurement for $\Gamma(f_2^{\prime}(1525) \to K^+K^-)$, one gets a value for the strong coupling of $3.80 \pm 0.16$ (with the incorrect dimension) that is around 5 times smaller than ours and therefore affecting the estimation of the branching fraction obtained in~\cite{Mohammadi}.

Both in the NR and R contributions, the decay width is parametrized in terms of the three-body phase space~\cite{PDG}
\begin{equation}
\Gamma(B_s^0 \to J/\psi K^+K^-)_{\rm NR(R)} = \dfrac{1}{32(2\pi)^{3} m_{B_s}^{3}}  \int_{t^{-}}^{t^{+}}  dt \int_{s^{-}}^{s^{+}} ds \ |\overline{\mathcal{M}}_{\rm NR(R)}|^{2},
\end{equation}

\noindent where $|\overline{\mathcal{M}}_{\rm NR(R)} |^{2}$ is the NR (R) spin-averaged squared amplitude\footnote{Their explicit expressions are provided in appendix~\ref{appB}.}. The integration limits are given by $t^{-} = 4 m_{K}^2$, $t^{+} = (m_{B_s} - m_{J/\psi})^{2}$ and
\begin{eqnarray}
s^{\pm}(t) &=& m_{B_s}^2 + m_K^2 - \dfrac{1}{2 t} \Big[t (t + m_{B_s}^2 - m_{J/\psi}^2)  \mp \lambda_t^{1/2} (t^2 - 4t m_{K}^2)^{1/2} \ \Big] .
\end{eqnarray}

\begin{figure}[!t]
\includegraphics[scale=0.33]{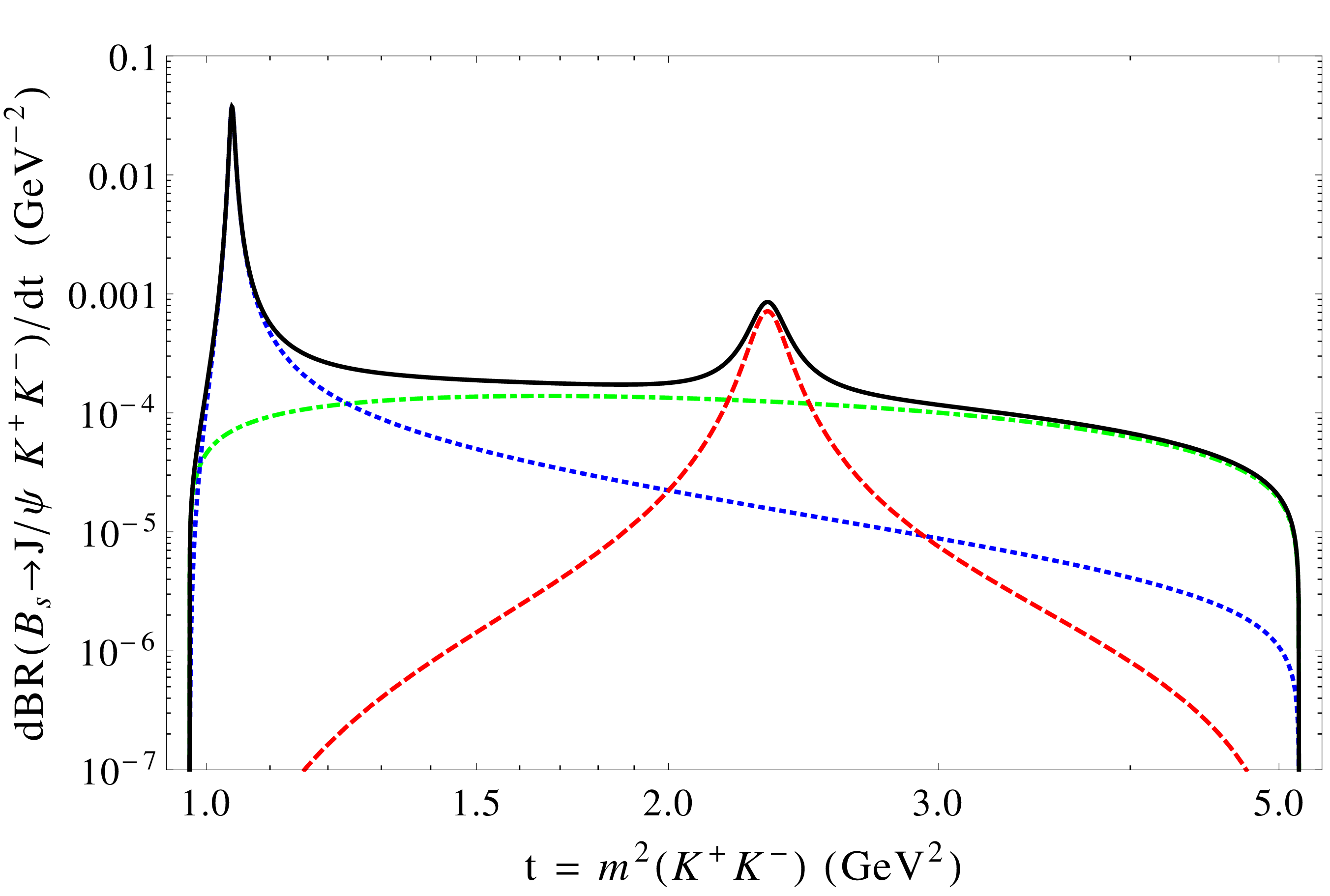} \ 
\includegraphics[scale=0.48]{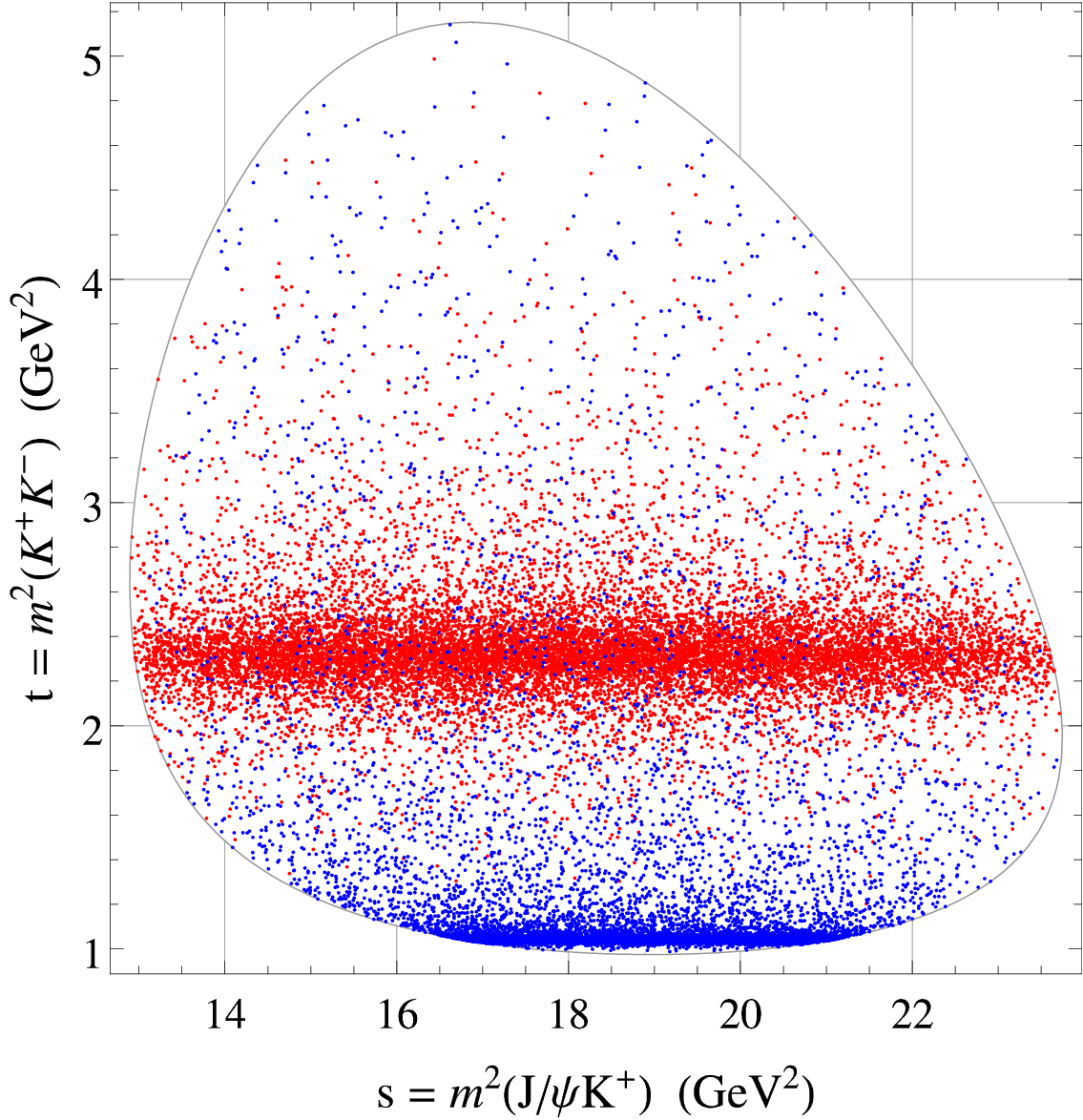}
\caption{\small [Left] Differential branching ratio of $B_s^0 \to J/\psi K^+K^-$ as function of $t=m^2(K^+K^-)$. The blue (dotted) and red (dashed) curves denote the contributions of resonances $V=\phi(1020)$ and $T=f_2^{\prime}(1525)$, respectively, while the NR contribution is represented by the green (dot-dash) curve. The black (solid) curve denotes the total contribution. [Right] Dalitz plot of $B_s^0 \to J/\psi K^+K^-$, where the horizontal blue and red bands represent the $\phi(1020)$ and $f_2^{\prime}(1525)$ resonances, respectively.}
\label{InvariantMass}
\end{figure}

\noindent with $\lambda_t = \lambda(t,m_{B_s}^2,m_{J/\psi}^2)$. In Figure~\ref{InvariantMass}[Left] we plot the differential branching ratio of $B_s^0 \to J/\psi K^+K^-$ as function of the invariant mass $m^2(K^+K^-)$. The black (solid) curve denotes the total contribution, while individual terms are given by the blue (dotted) curve for $V=\phi(1020)$,  red (dashed) curve for $T=f_2^{\prime}(1525)$, and NR contribution is represented by the green (dot-dash) curve.
As expected, the largest contribution is given by $\phi(1020)$ component, which it is clearly exhibited by the peak in Figure~\ref{InvariantMass}[Left], followed by  the $f_2^{\prime}(1525)$ component. There is also a sizeable contribution from NR term, which is dominated by the form factors $w_\pm$ with a negligible contribution from $h$. Comparing with the $m^2(K^+K^-)$ distributions obtained by LHCb (Figs. 15 and 17 of Ref.~\cite{LHCb:2013}), our distribution for the resonances agrees fairly well, showing a similar behavior. For the NR component, our distribution exhibits a different behavior to the LHCb, this is because a linear function has been used in the experimental analysis to describe the $K^+K^-$ mass~\cite{LHCb:2012,LHCb:2013}. As we will show below, this difference will turn out in a bigger estimation on the NR contribution than one reported by LHCb. 

As a complementary analysis, we perform the Dalitz plot of the process as shown in Figure~\ref{InvariantMass}[Right]. By using a Monte-Carlo simulation, we generate points $(s,t)$ over the phase space of $B_s^0\to J/\psi K^+K^-$ decay, with $s = m^2\left(J/\psi K^+\right)$ and $t =m^2\left(K^+K^-\right)$ the invariant masses. If the generated point $(s,t)$ fulfills the Cayley condition~\cite{Kajantie},  
\[G(t,s,m^2_{B_s},m^2_{K^+},m^2_{K^-},m^2_{J/\psi})\leq 0 , \] 

\noindent where $G$ is the Gram determinant~\cite{Kajantie}, we plot the point; otherwise we reject the point and select a new one until we get the Dalitz plot. The horizontal blue and red bands result from the $\phi(1020)$ and $f_2^{\prime}(1525)$ resonances, respectively. The obtained Dalitz plot is in accordance with the distribution obtained by LHCb (Fig. 6 of Ref.~\cite{LHCb:2013}).

The values of the different contributions to the total branching fraction of $B_s^{0} \to J/\psi K^+ K^-$ are summarized in Table~\ref{finalresult}. The error ranges are determined by the uncertainties on the above  couplings and then summed in quadrature. We predict a branching fraction of 
\begin{equation}
{\rm BR}(B_s^{0} \to J/\psi K^+ K^-) = (9.3{}^{+1.3}_{-1.1})\times 10^{-4},
\end{equation}

\noindent that is in agreement with experimental measurements reported by LHCb~\cite{LHCb:2013} and Belle~\cite{Belle:2013} (see Table  \ref{ExpResults}). Compared with the  previous theoretical estimation of $(10.3 \pm 0.9)\times 10^{-4}$~\cite{Mohammadi}, our result is consistent as well. However, in this previous work is unclear how much is the contribution both the NR and R components~\cite{Mohammadi}. In the present study a more detailed analysis on these contributions to the $m^2(K^+K^-)$ distribution and Dalitz plot is provided, thus extending the previous one~\cite{Mohammadi}. Moreover, keeping in mind that the value of the strong coupling constant $g_{T K^+K^-}$ was badly estimated (as it was above discussed), the theoretical value of the branching fraction obtained in~\cite{Mohammadi} can be incorrect.


Finally, let us mention that there are some works where only the $S$-wave contribution of the $K^+K^-$ spectrum of $B_s^{0} \to J/\psi K^+ K^-$ was estimated to be around $\sim 1.7 \%$~\cite{Oset:2015} and $\sim 1.1 \%$~\cite{Daub:2016}, while the contributions from $\phi(1020)$ and $f_2^{\prime}(1525)$ (as well as NR contribution) were not addressed in~\cite{Oset:2015,Daub:2016}. Furthermore, the authors of Ref.~\cite{Oset:2015} have estimated the ratio of branching fractions
\beq
\frac{{\rm BR}(B_s^{0} \to J/\psi K^+ K^-) }{{\rm BR}(B_s^{0} \to J/\psi \phi(1020)) } = (4.4 \pm 0.7)\times 10^{-2}.
\eeq

\noindent that is compatible within errors	with the experiment~\cite{Oset:2015}.

\begin{table}[!t]
\centering
\renewcommand{\arraystretch}{1.2}
\renewcommand{\arrayrulewidth}{0.8pt}
\caption{\small Values of the different contributions to the total branching fraction ($\times 10^{-4}$) of $B_s^{0} \to J/\psi K^+ K^-$.}
\begin{tabular}{cccc}
\hline\hline
  & \multicolumn{2}{c}{Resonant} &  \\
\cline{2-3}
Non-resonant & $V=\phi(1020)$ & $T=f_2^{\prime}(1525)$ & \textbf{Total} \\
\hline
$1.9 \pm 0.1$ & $5.6 \pm 0.7 $ & $0.8 {}^{+1.1}_{-0.8}$ & $9.3 {}^{+1.3}_{-1.1}$ \\
\hline\hline
\end{tabular} \label{finalresult}
\end{table}

\section{Concluding remarks} \label{conclusions}

Motivated by the phenomenological importance of nonleptonic charmonium $B_s$ decays, in this work we have carried out a reanalysis of the $B_s^0 \to J/\psi f_2^\prime(1525)$ and $B_s^0 \to J/\psi K^+K^-$ decays. Within the framework of the factorization a\-ppro\-ach and using the perturbative QCD for the evaluation of the relevant form factors, we have obtained a branching fraction for  the two-body channel of BR$(B_s^{0} \to J/\psi f_2^{\prime}(1525)) = (1.6 {}^{+0.9}_{-0.7})\times 10^{-4}$ which is in agreement with the experimental values reported by LHCb~\cite{LHCb:2013} and Belle~\cite{Belle:2013} Collaborations.
In addition, the polarization fractions associated with this vector-tensor mode have been studied for the first time. We found that the  fractions $f_T$ and $f_L$ are roughly equal, implying $f_T/f_L(J/\psi f_2^{\prime}) \sim 1$. This result is in agreement with theoretical prediction~\cite{Colangelo:2011,Wang:2014} and experimental measurement of the longitudinal polarization fraction obtained for the $B_s^{0} \to J/\psi \phi(1020)$ mode~\cite{LHCb:2011}.  Moreover, this is also in accordance with the fit fractions in the helicity basis obtained by the LHCb in the amplitude analysis of the $B_s^0 \to J/\psi K^+K^-$ decay for the resonance $f_2^{\prime}(1525)$~\cite{LHCb:2013}. 

Concerning the three-body mode $B_s^0 \to J/\psi K^+K^-$,  we have calculated both non-resonant and resonant contributions, and a detailed analysis of the $m^2(K^+K^-)$ distributions and Dalitz plot have been performed.  The non-resonant part has been described by the heavy meson chiral perturbation theory. For the resonant part, the contributions of the intermediate vector $\phi(1020)$ and tensor $f_2^{\prime}(1525)$ mesons have been taken into account by means of the Breit-Wigner resonance formalism. It is found that  the largest contribution is given by $\phi(1020)$ followed by  $f_2^{\prime}(1525)$, with a sizeable non-resonant contribution that agrees fairly well with the data~\cite{LHCb:2013}. The overall result of the branching fraction BR$(B_s^{0} \to J/\psi K^+ K^-) = (9.3 {}^{+1.3}_{-1.1})\times 10^{-4}$ is also in satisfactory agreement with the experimental data reported by LHCb~\cite{LHCb:2013} and Belle~\cite{Belle:2013}. 

\acknowledgments

The author N. Quintero acknowledges the support from Direcci\'{o}n General de Investigaciones - Universidad Santiago de Cali. The work of C. A. Morales,  C. E. Vera,  and A. Villalba has been supported by Comit\'{e} Central de Investigaciones - Universidad del Tolima under Project No. 330115. We are grateful to Carlos A. Ram\'{i}rez and Jos\'{e} Herman Mu\~{n}oz for reading the manuscript and providing suggestions. We are also indebted to Sheldon Stone, Liming Zhang, Diego Milan\'{e}s, and Alberto C. dos Reis for very helpful comments.


\appendix

\section{$B_s \to T$ form factors}  \label{appA}

The polarization of a generic tensor meson ($J^P = 2^+$) can be specified by a symmetric and traceless tensor $\tilde{\epsilon}^{\mu\nu}$ which satisfies the following properties~\cite{Wang11,Chen:2007,Datta08},
\bea
\tilde{\epsilon}^{\mu\nu}(p_T,\sigma) &=& \tilde{\epsilon}^{\nu\mu}(p_T,\sigma), \nonumber \\
\tilde{\epsilon}^{\mu\nu}(p_T,\sigma) p_{T\nu} &=& \tilde{\epsilon}^{\mu\nu}(p_T,\sigma) p_{T \mu} = 0, \nonumber
\eea

\noindent and $\tilde{\epsilon}^{\mu\nu}(p_T,\sigma) g_{\mu\nu} =0$, with $p_T$ and $\sigma$ the momentum and helicity of the $T$ meson. The states of a massive spin-2 particle can be constructed in terms of the spin-1 states as~\cite{Chen:2007}
\bea
\tilde{\epsilon}^{\mu\nu}(\pm 2) &=& e^\mu(\pm 1)e^\nu(\pm 1) , \nonumber \\
\tilde{\epsilon}^{\mu\nu}(\pm 1) &=& \dfrac{1}{\sqrt{2}}[e^\mu(\pm 1)e^\nu(0) + e^\nu(\pm 1)e^\mu(0)], \\
\tilde{\epsilon}^{\mu\nu}(0) &=& \dfrac{1}{\sqrt{6}}[e^\mu(+ 1)e^\nu(- 1) + e^\nu(- 1)e^\mu(+ 1)] + \sqrt{\dfrac{2}{3}} e^\mu(0)e^\nu(0) , \nonumber
\eea

\noindent with $e^\mu(0, \pm 1)$ denoting the polarization vectors of a massive vector state moving along the $z$ axis with the explicit structure~\cite{Chen:2007}
\bea
e^\mu(0) &=& \dfrac{1}{m_T} (|\vec{p}_T|,0,0,E_T) , \\
e^\mu(\pm 1) &=& \dfrac{1}{\sqrt{2}} (0,\mp 1,-i,0) ,
\eea

\noindent where $m_T$ and $|\vec{p}_T|$ ($E_T$) are the mass and the three-momentum magnitude (energy) of the $T$ meson in the $B_s$ rest frame, respectively. Defining the new polarization vector~\cite{Wang11,Chen:2007,Datta08,Cheng:2010}
\beq 
\epsilon_T^\mu = \tilde{\epsilon}^{\mu\nu} P_\nu /m_{B_s},
\eeq

\noindent which satisfies 
\bea
\epsilon_T^\mu(\pm 2) &=& 0 , \nonumber \\ 
\epsilon_T^\mu(\pm 1) &=& \dfrac{1}{\sqrt{2}} \Big(e(0).\dfrac{P}{m_{B_s}} \Big) \ e^\mu(\pm 1),  \\ 
\epsilon_T^\mu(\pm 0) &=& \sqrt{\dfrac{2}{3}} \Big(e(0).\dfrac{P}{m_{B_s}} \Big) \ e^\mu(0) , \nonumber 
\eea

\noindent with $e(0).P/m_{B_s} = |\vec{p}_T| / m_T$ and $P$ the $B_s$ meson momentum. We can see that although the tensor meson contains 5 spin degrees of freedom, only $\sigma = 0$ and $\pm 1$
give nonzero contributions. As a consequence the parametrization of the $B_s \to T$ form factors is analogous to the $B_s \to V$ case except that the $\epsilon^{\mu}_V$ is replaced by $\epsilon^{\mu}_{T}$.

In the Isgur-Scora-Grinstein-Wise (ISGW) model~\cite{ISGW}, the general expression for the $B_s \to T$ transition is parametrized as
\begin{eqnarray}
\langle T(p_T,\tilde{\epsilon})|\bar{s}\gamma_\mu b |B_s(P) \rangle &=& i h(q^2) \varepsilon_{\mu\nu\rho\sigma} \tilde{\epsilon}^{\ast \nu\alpha} \ P_{\alpha} (P+p_{T})^{\rho} q^{\sigma}, \nonumber\\
\langle T(p_T,\tilde{\epsilon})|\bar{s}\gamma_\mu\gamma_5 b|B_s(P) \rangle &=& \tilde{\epsilon}^{\ast}_{\alpha\beta} \ P^{\alpha}P^{\beta} [ b_{+}(q^2)(P +p_{T})_{\mu} + b_{-}(q^2) q_{\mu} ] + k(q^2) \tilde{\epsilon}^{\ast}_{\mu\nu} P^{\nu},  
\end{eqnarray}

\noindent where $q_{\mu} =(P-p_T)_{\mu}$ and $h, k, b_{\pm}$ are the form factors ($k$ is dimensionless and $h, b_{\pm}$ have dimension of GeV$^{-2}$) evaluated at the squared transfer momentum $q^2$. This set of form factors are related to the set $V^{B_sT}$ and $A^{B_sT}_{0,1,2}$ via~\cite{Cheng:2010}
\begin{eqnarray}
V^{B_s T}(q^2) &=& m_{B_s} (m_{B_s} + m_T) h(q^2) ,\nonumber \\
A^{B_s T}_1(q^2) &=&  \frac{m_{B_s} }{(m_{B_s} + m_T)} k(q^2), \\
A^{B_s T}_2(q^2) &=& - m_{B_s} (m_{B_s} + m_T) b_{+}(q^2), 	\nonumber \\
A^{B_s T}_0(q^2) &=& \frac{m_{B_s}}{2 m_T} [ k(q^2) + (m_{B_s}^2 - m_T^2) b_+(q^2) -t  b_{-}(q^2)]. \nonumber 
\end{eqnarray}

\section{Squared amplitudes}  \label{appB}
We collect in this appendix the non-resonant (NR) and resonant (R) spin-averaged squared amplitudes of the $B_s^0 \to J/\psi K^+K^-$ decay discussed in section~\ref{3body}. For NR contribution we have
\begin{align} \label{AmplitudNR}
|\overline{\mathcal{M}}_{\rm NR}|^2 = |\xi|^2
\Big[k_1(s,t) [\omega_+(s)]^2 + k_2(s,t)[\omega_-(s)]^2 + k_3(s,t)\omega_+(s)\omega_-(s) + k_4(s,t) [h(s,t)]^2 \Big],
\end{align}

\noindent where $\xi = i G_F V_{cb} V_{cs}^{*} \tilde{a}_{\rm eff} f_{J/\psi}m_{J/\psi}/\sqrt{2}$ and the kinematic factors $k_i(s,t)$ ($i=1,2,3,4$) are given by
\begin{align}
k_1(s,t)&=\frac{\lambda_t}{4m_{J/\psi}^2} , \\
k_2(s,t)&=\frac{1}{4m_{J/\psi}^2}\Big[m_{J/\psi}^4+2m_{J/\psi}^2\left(m_{B_s}^2-6m_K^2-2 s+t\right)+\left(2s+t-m_{B_s}^2-2m_K	^2\right)^2 \Big] , \\
k_3(s,t)&=\frac{1}{2m_{J/\psi}^2}\Big[m_{J/\psi}^4+\left(m_{B_s}^2-t\right)\left(2s+t-m_{B_s}^2-2m_K^2\right)-2m_{J/\psi}^2 (s-m_K^2) \Big] , \\
k_4(s,t)&= m_{J/\psi}^2  \Big[ t \big( s (m_{B_s}^2 + m_{J/\psi}^2 + 2 m_K^2) + (m_{J/\psi}^2 - m_K^2)(m_K^2 - m_{B_s}^2)  - s^2 \big) \nonumber \\
 & \ \ \ - m_K^2 (m_{J/\psi}^2 - m_{B_s}^2 )^2  - s t^2\Big],
\end{align}
 
\noindent with $m_K = m_{K^\pm}$ and $\lambda_t = \lambda(t,m_{B_s}^2,m_{J/\psi}^2)$. Let us notice that interference terms between $h$ and $w_\pm$ vanish. These kinematic factors are function of $s = m^2(J/\psi K^+)$, $t = m^2(K^+K^-)$ and the masses of mesons involved. 

The R contribution from $V=\phi(1020)$ reads as
\begin{align}
|\overline{\mathcal{M}}_V|^2 =& 
|\xi|^2 g^2_{VK^+K^-} c_{0V}(t)  \Big[c_{1V}(t)[A^{B_sV}_1(t)]^2+ c_{2V}(t)[A^{B_sV}_2(t)]^2  \nonumber \\
& + c_{3V}(t)[V^{B_sV}(t)]^2 + c_{4V}(t)A^{B_sV}_1(t) A^{B_sV}_2(t) \Big] ,
\end{align}

\noindent where $c_{0V}(t) = (t - 4 m_K^2) \vert{\rm BW}_{V}(t)\vert^2$ contains the information of the BW function [Eq.~\eqref{BW}] and $c_{iV}(t)$ ($i=1,2,3,4$) are kinematic factors defined by
\begin{align}
c_{1V}(t)&=\frac{(m_{B_s}+m_V)^2}{4tm_{J/\psi}^2} \big(\lambda_t + 12 m_{J/\psi}^2t\big), \\
c_{2V}(t)&=\frac{\lambda_t^2}{4tm_{J/\psi}^2 (m_{B_s}+m_V)^2}, \\ 
c_{3V}(t)&=\frac{2\lambda_t}{(m_{B_s}+m_V)^2}, \\
c_{4V}(t)&=\frac{\lambda_t}{2tm_{J/\psi}^2} (t-m_{B_s}^2+m_{J/\psi}^2).
\end{align}

\noindent As for the resonance $T=f_2^\prime(1525)$, we have
\begin{align}
|\overline{\mathcal{M}}_T|^2 =& 
|\xi|^2 g^2_{TK^+K^-} c_{0T}(t)  \Big[c_{1T}(t)[A^{B_sT}_1(t)]^2+ c_{2T}(t)[A^{B_sT}_2(t)]^2  \nonumber \\
& + c_{3T}(t)[V^{B_sT}(t)]^2 + c_{4T}(t)A^{B_sT}_1(t) A^{B_sT}_2(t) \Big] ,
\end{align}

\noindent where $c_{0T}(t) = (t - 4 m_K^2)^2 \vert{\rm BW}_{T}(t)\vert^2 / 24$ similarly contains the information of the BW function and the other $c_{iT}(t)$ ($i=1,2,3,4$) are given by
\begin{align}
c_{1T}(t)&=\lambda_t^2 /4t , \\ 
c_{2T}(t)&=\frac{\lambda_t}{24 t^2 m_{J/\psi}^2} \big(\lambda_t+10 m_{J/\psi}^2t\big),\\
c_{3T}(t)&= \frac{\lambda_t^3}{24 t^2 m_{J/\psi}^2}, \\
c_{4T}(t)&=\frac{\lambda_t^2}{12t^2m_{J/\psi}^2} (m_{B_s}^2- m_{J/\psi}^2 - t).
\end{align}


\end{document}